\documentclass[smallextended]{svjour3}       
\usepackage{cancel}
\usepackage{tikz}
\usepackage{pgfplots}
\usepackage{tikz-3dplot}
\usepackage{caption}
\usepackage{subcaption}
\tdplotsetmaincoords{60}{135}
\pgfplotsset{compat=newest}
\usepackage{graphicx}
\usetikzlibrary{calc,patterns,decorations.pathmorphing,decorations.markings, quotes,angles}
\usepackage{tikz-3dplot}

\makeatletter
\tikzset{
    dot diameter/.store in=\dot@diameter,
    dot diameter=3pt,
    dot spacing/.store in=\dot@spacing,
    dot spacing=10pt,
    dots/.style={
        line width=\dot@diameter,
        line cap=round,
        dash pattern=on 0pt off \dot@spacing
    }
}
\makeatother

\smartqed  
\usepackage[latin1]{inputenc}
\usepackage{graphicx}
\usepackage{tikz}
\usetikzlibrary{calc,patterns,decorations.pathmorphing,decorations.markings, quotes,angles}
\usepackage{tikz-3dplot}
\usepackage{amsmath,amsfonts,amssymb,amscd}
\usepackage{wasysym}
\usepackage{wrapfig}
\usepackage[colorlinks,linkcolor=blue,citecolor=blue,urlcolor=blue]{hyperref}
\usepackage{natbib}
\setcitestyle{round}

\usepackage{tikz}
\usetikzlibrary{calc,patterns,decorations.pathmorphing,decorations.markings}

\newcommand{\erm}{{\rm e} }

\usepackage[graphicx]{realboxes}

\newcommand{\sm}[1]{{\scriptscriptstyle#1}}

\newcommand{\Tr}{{\rm \!\ Tr\!\ } }

\newcommand{\Id}{{\bf 1}  }

\newcommand{\K}{\mathrm{K}}

\newcommand{\R}{\mathbb{R}}

\newcommand{\gm}{\gamma}

\newcommand{\mat}[1]{\mathbf{#1}}

\newcommand{\ov}{\overline}

\newcommand{\Ioa}{ {\,\rm I}_{\circ \alpha}}

\newcommand{\Io}{ {\,\rm I}_\circ}

\def\Z{{\mathbb Z}}

\def\R{{\mathbb R}}

\begin{document}

\title{Tidal Evolution and Spin-Orbit Dynamics: The Critical Role of Rheology.}



\author{    Clodoaldo Ragazzo    \and
           Lucas Ruiz dos Santos 
}

 \authorrunning{ Clodoaldo Ragazzo    \and  Lucas Ruiz} 

\institute{C. Ragazzo  (ORCID 0000-0002-4277-4173)\at
 Instituto de Matem\'{a}tica e Estat\'{i}stica, Universidade de S\~{a}o Paulo, 05508-090 S\~{a}o Paulo, SP, Brazil
 \\ \email{ragazzo@usp.br} \\
 Partially supported by FAPESP grant 2016/25053-8.
\and
L. S. Ruiz   (ORCID 0000-0002-5705-5278) \at
Instituto de Matem\'{a}tica e Computa\c{c}\~{a}o, Universidade Federal de Itajub\'{a}, 37500-903 Itajub\'{a}, MG, Brazil
\\ \email{lucasruiz@unifei.edu.br}
}


\maketitle

 \begin{abstract}

   This study analyzes secular dynamics using averaged equations that detail tidal effects on the motion of two extended bodies in Keplerian orbits.
   It introduces formulas for energy dissipation within each body of a binary system.
   The equations, particularly in contexts like the Sun-Mercury system, can be delineated into a fast-slow system.  A significant contribution of this work is the demonstration of the crucial role
   complex rheological models play in the capture by spin-orbit resonances.
   This is particularly evident in the notable enlargement of the basin of attraction for Mercury's current state when transitioning from a single characteristic time rheology to a dual characteristic time model, under the constraint that both models comply with the same estimate of the complex Love number at orbital frequency. The study also underscores the importance of Mercury's elastic rigidity on secular timescales.

 \end{abstract}

\section{Introduction}
\label{intro}

Newton formulated the law of gravitation and concluded that the motion of the centers
of mass of spherical bodies is equivalent to that of point masses. The solution to the resulting two-body problem,
as obtained by Newton, forms the backbone of all subsequent developments in celestial mechanics.
Notably, while planets and major satellites are almost spherical, even slight deformations
caused by spin and tidal forces can significantly influence their rotation and orbits.

Building on the foundational works of Newton, Laplace, Thomson, Darwin, and others, a significant advancement in understanding tidal effects on celestial motion was made by Kaula \cite{kaula1964tidal}, who decomposed tidal forces harmonically in space and time for two bodies in a Keplerian orbit, using Love numbers to evaluate changes in orbital elements. Recent updates to Kaula's theory are available in \cite{boue2019tidal}, with additional insights in \cite{efroimsky2012bodily}. Over the last 70 years, extensive research has focused on these tidal effects. Notable contributions include Ferraz-Mello's \cite{ferraz2013tidal} model, which builds on Darwin's theory with non-spherical hydrostatic states, detailed further in \cite{folonier2018tidal}, \cite{ferraz2020tidal}, and summarized in \cite{ferraz2019planetary}, \cite{ferraz2021tides}. Various studies, including those by \cite{goldreich1966final}, \cite{singer1968origin}, \cite{alexander1973weak}, \cite{mignard1979evolution}, \cite{hut1981tidal}, \cite{makarov2013no}, \cite{correia2014deformation}, \cite{ferraz2015tidal}, and \cite{boue2016complete}, have explored deformation equations averaged over orbital motion, particularly in low and high-viscosity scenarios.
The averaged equations used in this work are identical to those presented in \cite{correia2022tidal}.

In this paper we study the secular-planar dynamics of two extended bodies. We make the following assumptions:
\begin{itemize}
\item[$1)$] The  two bodies are  deformable, nearly spherical at all times; 
\item[$2)$] The spins (or rotation vectors) of the deformable bodies remain perpendicular to the orbital plane.
\item[3)] The bodies are: radially stratified, each body layer is homogeneous and
has a linear visco-elastic rheology,  see e.g.  \cite{sabadini2016global}.
Fluid layers, if present, must be sufficiently coupled to
the adjacent layers such that the rotation of each layer remains close to the average rotation of
the body.\end{itemize}

In this paper, we  crucially use that, from the perspective of gravitation,  the rheology of
a body with a finite number of homogeneous layers 
is equivalent to that of a homogeneous body with a sufficiently  more complex
rheology \cite{gevorgyan2023equivalence}.

The aim of this study is to analyze the secular dynamics arising from the averaged equations. The primary novelty of this work lies in demonstrating the importance of using complex rheological models in the capture by spin-orbit resonances. More specifically, it highlights the significant enlargement of the basin of attraction of Mercury's current state when transitioning from a rheology with one characteristic time to one with two characteristic times. Both rheologies are constrained to satisfy the same estimate of the complex Love number at orbital frequency.

In the next section, we present the main results of the paper. The final section provides several mathematical details involved in deriving these results.

  \section{Main results.}

  The foundational equations for the orbit and rotation of an extended body are well-established in the literature. Various equations detailing the deformation of extended bodies exist. In a companion paper \cite{rr2023a}, we averaged the equations provided in \cite{rr2017} with respect to the orbital motion, excluding the term accounting for the inertia of deformations as discussed in \cite{correia2018effects}. These equations are applicable to any rheological model. We obtained essentially the same averaged equations as those presented in \cite{correia2022tidal}. The only difference is the inclusion of a centrifugal deformation term, which is not relevant in the planar case. It is important to emphasize that the averaged equations in \cite{correia2022tidal} are more general as they do not necessitate the spins of the bodies to be perpendicular to the orbital plane.

Let $m_{\alpha}$ and $m$ represent the masses of two celestial bodies,
which could be a planet and a star, or a planet and a satellite, etc. We name the bodies
such that the ``$\alpha$ body''  is the largest  $  m_\alpha\ge m$.
The quantities of the large  body will always be labeled  with an index $\alpha$ and those of
the small body will have no label.

We assume that both bodies  are almost spherical,  deformable, and the deformations are volume preserving.
In this situation the mean moments of inertia, denoted as   $\Io$ and ${\rm I}_{\circ \alpha}$,
remain  constant in time, a result attributed to Darwin
\cite{rochester1974changes}.

The motion of the two-body system is determined by three sets of equations: the equations for the
relative positions of the centers of mass (orbit), the equations for the rotation of each body about
their centers of mass (spin), and the equations for the deformation of each body.
In Section \ref{eqsec}, we present these fundamental equations.

\subsection{Rheology and Love numbers}

The rheology of a body determines the Love number $k_2(\sigma)$, where $\sigma$ is the angular
frequency of the tidal raising force. For a stratified body with a finite number of homegeneous
layers, the  Love number can be written  as \cite{sabadini2016global} (see also
\cite{gevorgyan2023equivalence} and  \cite{gev2021}):
\begin{equation}\begin{split}
         k_2(\sigma)& =k_\infty+ (k_\circ-k_\infty)
         \left(\frac{h_1}{1+i\tau_1\sigma}+\cdots+\frac{h_{n+1}}{1+i\tau_{n+1}\sigma}\right).\ \
\end{split}
        \label{k2voigt2}
\end{equation}
In these equations:
\begin{itemize}
\item[$\bullet$]
  $k_\infty:=\lim_{\sigma\to\infty}k_2(\sigma)$ is the asymptotic value of the Love number
  at high frequencies.
\item[$\bullet$]  $k_\circ:=k_2(0)$ is the Love number at frequency zero, also called
  secular Love number  \footnote{
    The softest possible body is one composed of a perfect fluid, which
    is held together solely by self-gravity. In this case, $k_\circ = k_f$,
    where $k_f$ is the fluid Love number. Let $R$ be the body volumetric radius and
    $R_I$ be the radius of inertia, defined as the radius of a homogeneous sphere of mass $m$ and moment of
inertia $\Io$.
Assuming that the density is non-decreasing towards the center, the following approximation 
(valid for $0.2 < \frac{\Io}{mR^2} \le 0.4$) holds 
\citep[Eqs. 1.2 and 1.8, Theorems 4.1 and 4.2]{ragazzo2020theory}:
\begin{equation}  
  k_f \approx \frac{3}{2}\left(\frac{R_I}{R}\right)^5 \quad \text{where:}\quad R_I^2 := \frac{5}{2} \frac{\Io}{m}\,.\label{clairaut}
  \end{equation} The maximum value of $R_I/R$ is one, achieved in a homogeneous body,
for which $k_f = \frac{3}{2}$. The approximation in equation (\ref{clairaut}) was also
proposed in \cite{consorzi2023relation}.}.
 \item[$\bullet$] $\tau_i$ is the $i^{th}$ characteristic time of the rheology.
  \item[$\bullet$] $h_i>0$ is the relative amplitude of the $i^{th}$ mode of the rheology, which
    is associated to the characteristic time $\tau_i$. The relative amplitudes add to one:
    \[h_1+h_2+\ldots h_{n+1}=1.\]
    \end{itemize}

\begin{figure}[h]
    \begin{center}
        \includegraphics[width=0.9\textwidth]{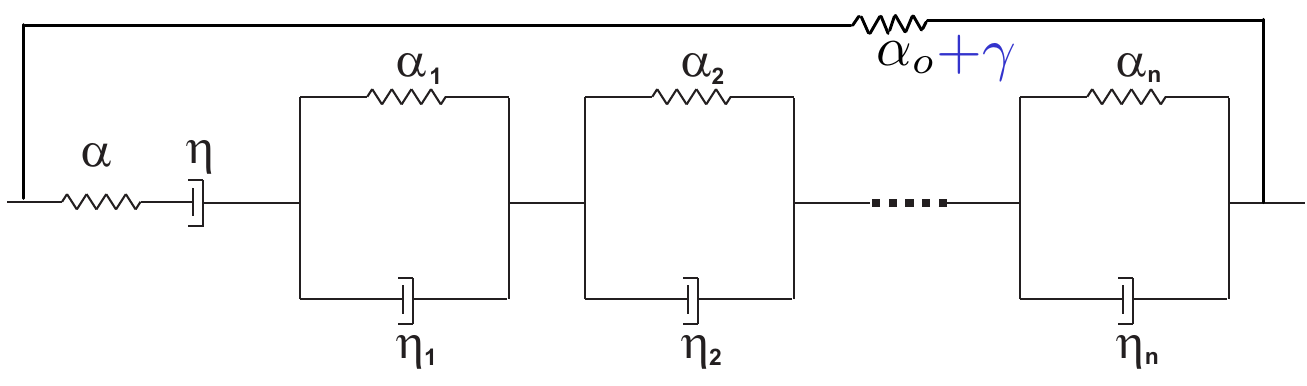}
        \caption{The generalized Voigt model. The constant $\gamma$ (blue)
          represents self-gravity rigidity.}
        \label{genvoi}
    \end{center}
\end{figure}

As discussed in \cite{gevorgyan2023equivalence}, the Love number in equation (\ref{k2voigt2})
corresponds to that of a homogeneous body with a generalized Voigt rheology, the spring-dashpot 
representation of which is given in Figure \ref{genvoi}.  

\begin{figure}[hptb!]
\centering
\begin{minipage}{0.5\textwidth}
\centering
\includegraphics[width=0.9\textwidth]{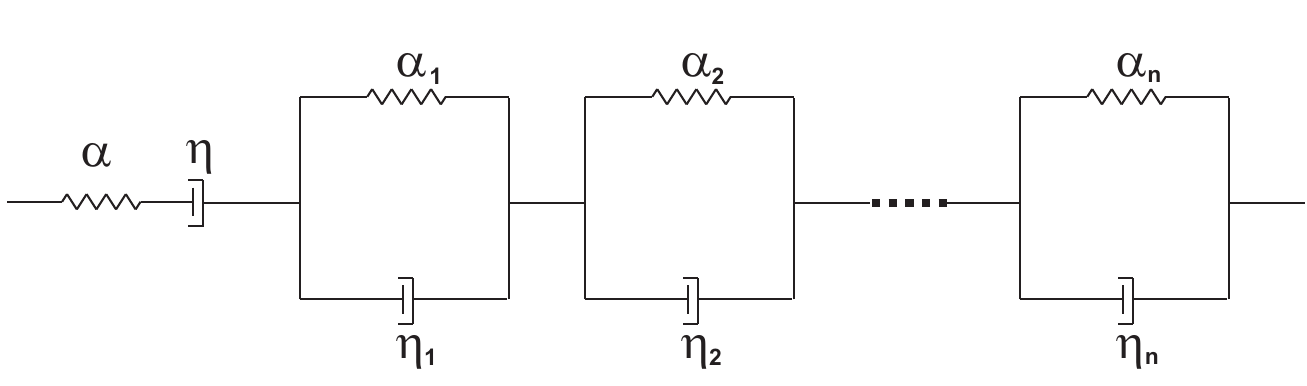}
\end{minipage}\hfill
\begin{minipage}{0.45\textwidth}
  \centering
  \includegraphics[width=0.9\textwidth]{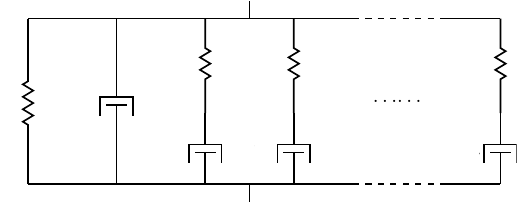}
\end{minipage}\hfill
\caption{
  LEFT: Generalized Voigt model in \cite{Bland} Figure 1.8.
  RIGHT: Generalized Maxwell model \cite{Bland} Figure 1.9.
 }
\label{fbland}
\end{figure}

All the simple rheological models used in the literature, such as: Maxwell, Kelvin-Voigt,
Standard Anelastic Solid (SAS), Burgers,  Bland's generalized Voigt  (Figure \ref{fbland} LEFT),  and
Bland's generalized Maxwell  (Figure \ref{fbland} RIGHT);
 are particular cases of the generalized Voigt rheology depicted in Figure
\ref{genvoi}.
The Andrade model can be approximated with arbitrary precision by a generalized Voigt model, as demonstrated in \cite{gev2020}.

\subsection{Passive deformation.}

The moment of inertia matrix $\mathbf{I}$ of a  body in an inertial reference frame
 can be written as:
\begin{equation}
  \mathbf{I}= \Io\big(\Id-\mathbf{b}\big).\label{Ib}
\end{equation}
where $\Id$ is the identity and $\mathbf{b}$ is a symmetric and traceless matrix.
We denote matrices and vectors in bold face. The matrix $\mat b$ is termed the deformation matrix. 
The deformation matrix of the $\alpha$-body is denoted as $\mat b_\alpha$. 

Since the bodies are nearly spherical and tidal deformations are minor, the variations in orbital
elements  occur gradually. As a result, the tidal forces can be approximated by those of point
masses undergoing
Keplerian motion. Within this framework, tidal forces can be harmonically decomposed both temporally,
using Hansen coefficients\footnote{
  If  $a$ is the semi-major axis, $e$ is the eccentricity, and $M$ is the mean anomaly, then
  \begin{equation}
 \left(\frac{r}{a}\right)^n\erm^{i m f}=\sum_{k=-\infty}^\infty X^{n,m}_k(e) \erm^{i k M}\,,
 \label{hanseneq}
 \end{equation}
 where: $n$, $m$, and $k$ are integers, and $X^{n,m}_k(e)$ is  the Hansen coefficient.},
and spatially, using spherical harmonics.

In Section \ref{passec},
we employ these decompositions to compute the ``passive deformation matrix'' in terms of
Love numbers, orbital eccentricity,  mean motion ($n=\dot M$), and the spin angular
velocities ($\omega$ and $\omega_\alpha$) of each body. The term ``passive'' refers to the fact
that this deformation is computed while neglecting its influence on altering the orbital elements and
the spin of each body.

\subsection{Energy and angular momentum.}

The energy function of a system of two rigid spherical bodies is
\begin{equation} 
  E_\circ:=-\frac{G m m_\alpha}{2a}+\Io\frac{\omega^2}{2}+\Ioa\frac{\omega_\alpha^2}{2}=
  - a_1n^{2/3}
  +\Io\frac{\omega^2}{2}+\Ioa\frac{\omega_\alpha^2}{2}
\,,
  \label{Eo}\end{equation}
where: $G$ is the gravitational constant,
$a$ is the semi-major axis, and
\begin{equation}
  a_1:=\frac{ mm_\alpha G^{2/3}}{ (m+m_\alpha)^{1/3}}\,.\label{a1}
  \end{equation}
The  angular momentum of a spherical body is $\ell_s=\Io \omega$ and the  orbital
angular momentum is
\begin{equation}
  \ell=\frac{mm_\alpha}{m+m_\alpha}a^2n\sqrt{1-e^2}=a_1 n^{-1/3}\sqrt{1-e^2}\,.\label{ell}
  \end{equation}
  The total angular momentum of  a system of two spherical bodies is
\begin{equation}
  \ell_{\scriptscriptstyle T}:=\ell+\ell_s+\ell_{s\alpha}.\label{ellt}
\end{equation}
We will assume that $\ell>0$ (or $n>0$) and $\ell_{\scriptscriptstyle T}>0$  for all time.

  The energy and angular momentum of a system of two slightly deformable bodies, which are
  given in Section
  \ref{eqsec}, are approximately given by the expressions of their spherical approximations.

  The total angular momentum is constant,   $ \dot \ell_{\scriptscriptstyle T}=0$, and
  within the Lagrangian formalism with dissipation function \cite{rr2017} the time derivative
  of the energy is given by the sum of the dissipation functions of each body
  $\cal D$ and $\cal D_\alpha$.

  In Section \ref{secav} we use the passive deformations to estimate the average dissipation of energy
  in each body. For the small body the result is:
  \begin{equation}\begin{split}
  \langle \mathcal{D}\rangle 
 &=- \frac{3}{8} \left(\frac{m_\alpha}{m+m_\alpha}\right)^2  \frac{n^4R^5}{G} \sum_{k=-\infty}^\infty\Bigg\{
 \frac{kn}{3}\Big(X^{-3,0}_k(e)\Big)^2 {\rm Im}\,  k_2(k n)
 \\
 &\ +(k n-2\omega)\Big(X^{-3,2}_k(e)\Big)^2 {\rm Im}\,  k_2(k n-2\omega)
 \Bigg\}.
 \end{split}\label{da1}
\end{equation}
A similar expression holds for the $\alpha$-body.

Since Im $k_2(-\sigma)=-$Im $k_2(\sigma)$ and Im $k_2(\sigma)<0$ for $\sigma>0$, then
$\langle \mathcal{D}\rangle\ge 0$. If $e=0$, then $X^{-3,0}_k(0)=0$ for  $k\in\Z$ and
and $X^{-3,2}_k(0)=0$ for $k\ne 2$. So $\langle \mathcal{D}\rangle=0$ if and only if $e=0$ and $n=\omega$.

The energy dissipated due to passive deformations must originate from the
motion of the spherical bodies that induce these passive deformations.
Consequently, we can infer
\begin{equation}\begin{split}
    \dot E_\circ&=
 -\frac{2}{3} a_1n^{-1/3}\dot n
  +\Io\omega\dot\omega+\Ioa\omega_\alpha\dot\omega_\alpha\\
    &=-2\langle\mathcal{D}\rangle-2\langle\mathcal{D_\alpha}\rangle. \end{split}\label{th1}
\end{equation}

\subsection{The average torque and the secular equation for
  the orbital elements}

For the small body, the average torque due to passive deformations is:
  \begin{equation}\begin{split}
      \langle {\rm T}\rangle &=
      -  \frac{3}{2} \left(\frac{m_\alpha}{m+m_\alpha}\right)^2  \frac{n^4R^5}{G}
       \bigg\{\sum\limits_{k=-\infty}^\infty
 \Big(X^{-3,2}_{k}\Big)^2{\rm Im} \, k_2(k n-2\omega)\bigg\}.
    \end{split} \label{avt1}
 \end{equation}  
A similar  expression holds for the $\alpha$-body.

The secular equation for the orbital elements can be readily derived from equations
(\ref{da1}), (\ref{th1}), (\ref{avt1}), and their counterparts for the $\alpha$-body:
\begin{equation}\begin{split}
     \dot \omega& =  \frac{1}{\Io}\langle {\rm T}\rangle,\\
   \dot \omega_\alpha & = \frac{1}{ \Ioa} \langle {\rm T}_\alpha\rangle,\\
    \dot n& =\frac{3 n^{1/3}}{2 a_1}\Big(
2\langle\mathcal{D}\rangle+2\langle\mathcal{D_\alpha}\rangle   +\omega \langle {\rm T}\rangle
 +\omega_\alpha \langle {\rm T}_\alpha\rangle\Big).
\end{split}\label{amain}
\end{equation}
The conservation of total angular momentum, combined with equations (\ref{ell}) and (\ref{ellt}), implies
that the eccentricity, which is featured on the right-hand side of equations (\ref{amain}),
can be expressed in terms of the state variables $\omega, \omega_\alpha, n$.

\subsection{Time scales and a simplification when $m\ll m_\alpha$.}

The despin rate is dependent on the imaginary parts of the Love numbers. The largest bodies
in the solar
system are fluid  (e.g., the Sun, Jupiter, etc.) and  have an imaginary part of the
Love number that is significantly smaller
than that of bodies with solid components. Therefore, it is reasonable
to assume in the subsequent equations
that either ${\rm Im}\, k_2$ and ${\rm Im}\, k_{2\alpha}$ are comparable
or ${\rm Im}\, k_2\gg {\rm Im}\, k_{2\alpha}$.

The ratio between the despin rates, assuming no spin-orbit resonances, is given by
\begin{equation}\frac{\dot \omega}{\dot \omega_\alpha}\approx \frac{\Ioa}{\Io}
  \frac{m_\alpha^2}{m^2}\frac{R^5}{R_\alpha^5}\frac{{\rm Im}\, k_2}{{\rm Im}\, k_{2\alpha}}
  \approx \frac{\rho_\alpha}{\rho}
  \frac{m_\alpha^2}{m^2}\frac{{\rm Im}\, k_2}{{\rm Im}\, k_{2\alpha}}\,,
\end{equation}
where $\rho$ represents the density of the body.
If $m_\alpha \gg m$, then $\dot \omega \gg \dot \omega_\alpha$, indicating that the despin rate
of the larger body is significantly slower than that of the smaller body. In such a case, as a first
approximation, we may assume  $\omega_\alpha=$constant and  equations (\ref{amain}) become:
\begin{equation}\begin{split}
     \dot \omega& =  \frac{1}{\Io}\langle {\rm T}\rangle,\\
    \dot n& =\frac{3 n^{1/3}}{2 a_1}\Big(
2\langle\mathcal{D}\rangle+2\langle\mathcal{D_\alpha}\rangle   +\omega \langle {\rm T}\rangle
\Big).
\end{split}\label{amain2}
\end{equation}

For example, for the Sun-Mercury
system, using
Im$k_{2\alpha}\approx 3.5\times 10^{-8}$ \cite{ogilvie2014tidal} and
Im$k_2=0.0051$ \cite{margot2018mercury}, we obtain $\frac{\dot \omega}{\dot \omega_\alpha}\approx 2.8\times 10^{17}$.

\subsection{A simplification when the $\alpha$ body
  is fluid (almost inviscid).}

The ratio of the energy dissipation rates in each body is given by
\begin{equation}\frac{\langle {\cal D}\rangle}{\langle {\cal D}_\alpha\rangle}\approx 
  \frac{m_\alpha^2}{m^2}\frac{R^5}{R_\alpha^5}\frac{{\rm Im}\, k_2}{{\rm Im}\, k_{2\alpha}}
  \approx 
  \frac{\rho_\alpha^2}{\rho^2}\frac{R_\alpha}{R}\frac{{\rm Im}\, k_2}{{\rm Im}\, k_{2\alpha}}.
  \label{eqLuc1}
\end{equation}
Equation \eqref{eqLuc1}, along with the equation for $\dot n$ in (\ref{amain2}), implies that depending on the imaginary parts
of the Love numbers, the larger body may play a more significant role than the smaller body in altering the orbital
elements \footnote{For the Earth-Moon system, using the Love number $k_{2\alpha} = 0.2817 - 0.02324 \,i$
for the Earth \cite[Table 3, semi-diurnal frequency]{rr2017}, and
${\rm Im} \, k_2 = 5.13 \times 10^{-4}$ for the Moon (orbital frequency)
\cite{fienga2019inpop19a}, we obtain
\begin{equation}
 \frac{\langle {\cal D}\rangle}{\langle {\cal D}_\alpha\rangle}\approx  \frac{m_\alpha^2  R^5 \, {\rm Im} \, k_2}{
  m^2  R_\alpha^5 \,{\rm Im} \, k_{2\alpha}} = 0.022. 
\end{equation}
This indicates that the Earth's influence on altering the orbital parameters
is greater than that of the Moon.
}.

However, when the $\alpha$ body consists of a low viscosity fluid and the smaller body includes a solid
part, then $\frac{{\rm Im}\, k_2}{{\rm Im}\, k_{2\alpha}} \gg 1$, and
$\frac{\langle {\cal D}\rangle}{\langle {\cal D}_\alpha\rangle}$ may also be significantly greater than one.
This scenario applies to the Sun-Mercury system, where
$\frac{\langle {\cal D}\rangle}{\langle {\cal D}_\alpha\rangle} \approx 2.8\times 10^6$. In such cases,
$\langle {\cal D}_\alpha\rangle$ can be neglected in comparison to $\langle {\cal D}\rangle$ in
the equation for $\dot n$ in (\ref{amain2}).

Therefore, for the Sun-Mercury system, we can employ the following approximation to equation
(\ref{amain}):
\begin{equation}\begin{split}
     \dot \omega &=  \frac{1}{\Io}\langle {\rm T}\rangle,\\
    \dot n &= \frac{3 n^{1/3}}{2 a_1}\Big(
2\langle\mathcal{D}\rangle  +\omega \langle {\rm T}\rangle
\Big)\, ,
\end{split}\label{amain3}
\end{equation}
where the state variables are $(\omega, n)$. This type of equation commonly appears
in the literature, as seen in references such as \cite{correia2014deformation}, \cite{gomes2019rotation},
and \cite{correia2022tidal}.

The  angular momentum, which remains constant along the motion, becomes simpler:
\begin{equation}
  \ell_{\scriptscriptstyle T}=a_1 n^{-1/3}\sqrt{1-e^2}+\Io \omega.
  \label{ellt2}
\end{equation}

Equations (\ref{amain3}) are equivalent to those
obtained  in  \cite{correia2022tidal} in the planar case.

\subsection{The dynamics when $\frac{\Io}{m a^2} \ll 1$.}

\label{secma}

In the remainder of this section, we consider the simplest
case where equation (\ref{amain3}) is applicable, with $m \ll m_\alpha$ and
${\rm Im}\, k_{2\alpha} \ll {\rm Im}\, k_2$.

The ratio
\begin{equation}\begin{split}
  \frac{\dot n}{\dot \omega}
 & = 
       \frac {\Io n^{4/3}}{ a_1}\frac{3\omega}{2n}\left\{1+{ \scriptstyle\frac{\frac{1}{2 \omega}
     \Bigg\{  \sum\limits_{k=-\infty}^\infty
 \frac{kn}{3}\Big(X^{-3,0}_k(e)\Big)^2 {\rm Im}\,  k_2(k n)+\ldots
 \Bigg\}}{\sum\limits_{k=-\infty}^\infty
       \Big(X^{-3,2}_{k}\Big)^2{\rm Im} \, k_2(k n-2\omega)}}\right\}\\
     \end{split}   \label{rationo}  
   \end{equation}
includes the factor
   \begin{equation}
   \frac{ \Io n^{4/3}}{ a_1}  = \frac{1}{m}\left(1+\frac{m}{m_\alpha}\right)
  \frac{\Io}{a^2}\approx
  \frac{\Io}{m a^2}\,.\label{a1in}
\end{equation}

We assume
 \begin{equation}
   \frac{ \Io n^{4/3}}{ a_1}  \approx
  \frac{\Io}{m a^2} \ll 1 \,.\label{a1in2}
\end{equation}
For the Sun-Mercury system,
\begin{equation} \frac { \Io n^{4/3}}{ a_1} =
   6.2 \times 10^{-10}\,.\label{smno}
  \end{equation}

\begin{figure}[ptb]
\begin{center}
  \includegraphics[scale=0.9]{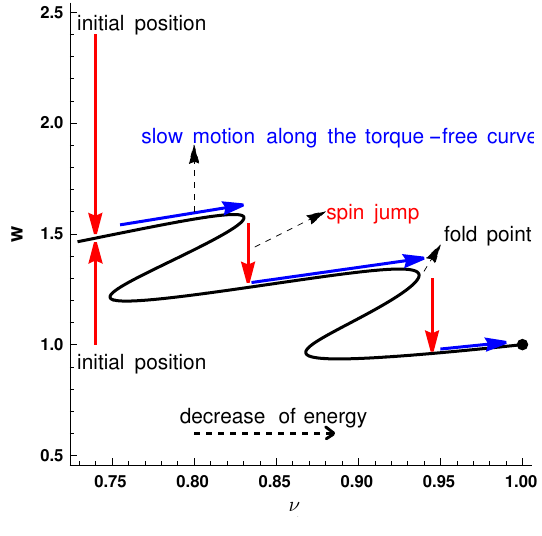}
\end{center}
\caption{Typical dynamics when $\frac{m a^2}{\Io} \gg 1$.
We use the nondimensional mean motion $\nu = \frac{\ell^3_{\scriptscriptstyle T}}{a^3_1} \, n$ and
the nondimensional spin angular velocity ${\rm w} = \frac{\ell^3_{\scriptscriptstyle T}}{a^3_1} \, \omega$
in the plot, as seen in equation (\ref{nuw}) and the previous paragraphs. 
The curve in black represents the torque-free curve $\cal C$: $\dot {\rm w} < 0$ above $\cal C$,
$\dot {\rm w} > 0$ below $\cal C$, and $\dot {\rm w} = 0$ on $\cal C$. An initial condition
above $\cal C$ converges rapidly to $\cal C$ with a decreasing spin velocity, while an initial
condition below $\cal C$ converges rapidly to $\cal C$ with an increasing spin velocity.
Upon reaching $\cal C$, the solution gradually progresses along $\cal C$.
The decrease in energy implies that $\nu$ increases along $\cal C$ (with $e$ decreasing).
The motion along $\cal C$ is sustainable up to a fold point. At this point, the decrease in energy
forces the solution to depart from $\cal C$, resulting in a spin jump. The solution is then drawn
towards another point on $\cal C$, representing a lower-order spin-orbit resonance.
Eventually, the solution approaches the synchronous
solution $(\nu, {\rm w}) \approx (1, 1)$. For a general discussion on the flow close to $\mathcal{C}$, see \cite{rr2023a}.
}
\label{fs}
\end{figure}

If $\left|\frac{\omega}{n}\right|$ is not far from one and  $(n,\omega)$ is not on
\begin{equation}
  {\cal C} := \bigg\{(n,\omega):\sum\limits_{k=-\infty}^\infty
       \Big(X^{-3,2}_{k}(e)\Big)^2{\rm Im} \, k_2(k n-2\omega)
  =0\bigg\}\quad \text{\bf (torque-free curve)}\,,
  \label{Ckv}
\end{equation} then 
$ \frac{\dot \omega}{\dot n} \gg 1$.
As a result, the spin-orbit dynamics is governed by a slow-fast
system of equations, where $n$ is the slow variable and $\omega$ is the fast variable.
The typical dynamics is described in the caption of Figure \ref{fs}.

Due to inequality (\ref{a1in2}), equation (\ref{ellt2}) can be simplified to
\begin{equation}
  \frac{\ell_{\scriptscriptstyle T}}{ n \Io} = \frac{a_1}{\Io n^{4/3}} \sqrt{1-e^2} + \frac{\omega}{n}
  \approx \frac{a_1}{\Io n^{4/3}} \sqrt{1-e^2}\,, \label{ellt3}
\end{equation}
where we have used that $\left|\frac{\omega}{n}\right|$ is not far from one and $e$ is not close
to one.
This leads to the approximation 
\begin{equation}
  e \approx \sqrt{1 - \left(\frac{\ell^3_{\scriptscriptstyle T}}{a^3_1 } n\right)^{2/3}}.\label{en}
\end{equation}

The constant $\frac{\ell^3_{\scriptscriptstyle T}}{a^3_1 }$ has the unit of time
(for the Sun-Mercury system $\frac{\ell^3_{\scriptscriptstyle T}}{a^3_1 }=13.1224$ days\footnote{
  $2 \pi\frac{\ell^3_{\scriptscriptstyle T}}{a^3_1 }=82.45$ days
  is approximately  the period of the
  longest  possible synchronous, $\omega=n$,  circular orbit of Mercury about the Sun.}).
We will use this constant to
define the nondimensional mean motion $\nu$ and spin angular velocity ${\rm w}$, which
are used in several figures:
\begin{equation}
    \nu := \frac{\ell^3_{\scriptscriptstyle T}}{a^3_1 } \, n \qquad
    {\rm w} := \frac{\ell^3_{\scriptscriptstyle T}}{a^3_1 } \, \omega.
    \label{nuw}
\end{equation}

\subsection{Characteristic times of terrestrial planets. }

\label{terr}

The imaginary part of the Love number of Mercury remains largely unknown, even at its orbital frequency
\cite{baland2017obliquity}, \cite{steinbrugge2018viscoelastic}. The most well-studied terrestrial bodies in are
the Earth and the Moon. It is reasonable to hypothesize that the rheology of Mercury could be
similar, at least qualitatively, to those of the Earth and Moon.

Thirteen characteristic times and amplitudes of a five-layer interior model for the Moon
\cite{matsuyama2016grail}
were computed in  \cite{gevorgyan2023equivalence}. This research demonstrated that the complex Love number
of the model can be effectively represented within the frequency range of interest by just
three modes. These modes have nondimensional amplitudes $h$ and characteristic times $\tau$ (in days)
(\cite{gevorgyan2023equivalence} Table 2 \footnote{
  The correspondence between the notation $(s_j,r_j)$ in \cite{gevorgyan2023equivalence} and that used here is:
  $s_j=-\tau_j^{-1}$, $r_j=h_j(k_\circ-k_\infty)$. The values of $h_1,h_2,h_3$ in (\ref{moon})
  are normalized such that $h_1+h_2+h_3=1$}):
\begin{equation}\begin{split}
  (h_1,\tau_1)&=\big(0.487, 12.2\, {\rm d}\big)\,,\\
  (h_2,\tau_2)&=\big(0.512, 18.3\, {\rm d}\big)\,,\\
  (h_3,\tau_3)&=\big(0.001, 8.3 \times 10^6\, {\rm d}\big).\\
\end{split}\label{moon}
\end{equation}

A five-layer interior model for the Earth is presented in Table 2-1 in \cite{sabadini2016global}. 
Nine relaxation times for this model, ranging from $2.6$ to $2 \times 10^5$ centuries, are provided 
in \cite[p.64]{sabadini2016global}, but without corresponding amplitudes.

These examples demonstrate that terrestrial bodies, such as Mercury:
\begin{itemize}
  \item[$\bullet$] May possess more than one relevant characteristic time,
  \item[$\bullet$] The relevant characteristic times can differ by orders of magnitude.
\end{itemize}

To investigate the importance of Mercury's rheology in the spin-orbit dynamics, we compare two different scenarios. In the first, 
Mercury is assumed to have a rheology with exactly one characteristic time $\tau$. 
In the second scenario,
Mercury is assumed to have a rheology with exactly two characteristic times $\tau_1$ and $\tau_2$, 
such that $\tau_1 \ll \tau_2$ and $\tau_2n_{mer} \gg 1$, where $n_{mer}$ is the orbital mean motion.

\subsection{Mercury's Love Number.}

      For Mercury, at the orbital frequency
      \begin{equation}
        n_{mer}= \frac{2 \pi}{87.969 \text{ days}}\,,
        \end{equation}
current estimates of the Love number fall within the range
$0.53 < {\rm Re}\, k_2 < 0.63$, with the preferred value being $k_2 = 0.569$ \cite{genova2019geodetic},
\cite{goossens2022evaluation}, and $0 < -{\rm Im}\, k_2 < 0.025$
\cite[p. 152]{baland2017obliquity}, \cite[p.2767 and 2769]{steinbrugge2018viscoelastic}.
We will assume ${\rm Re}\, k_2 = 0.57$ and $-{\rm Im}\, k_2 = 0.011$ as reference values
(quality factor $Q \approx -\frac{{\rm Re}\, k_2}{{\rm Im}\, k_2} \approx 52$).

Mercury is not in hydrostatic equilibrium: the expected hydrostatic flattening
is two orders of magnitude lower than the observed flattening. 
This discrepancy in hydrostaticity may be attributed to Mercury's despinning history
\cite{matsuyama2009gravity}. According to \cite{ragazzo2022librations} (Section 4.2), 
this suggests the presence of secular-elastic rigidity, represented by $\alpha_0 > 0$ in Figure \ref{genvoi},
in conjunction with self-gravity, such that
$k_\circ < k_f$, where $k_f$ denotes the fluid Love number.
Equation (\ref{clairaut}) implies that $k_f = 1.04$. 

The choice of $k_\circ$ within the range $[0.57, 1.04]$ is critical. One of the main results we present below -- the significant enlargement of the basin of attraction of Mercury's current state when transitioning from a rheology with one characteristic time to one with two characteristic times -- does not hold if $k_\circ = k_f$. This highlights the importance of Mercury's secular elastic rigidity in determining
its current state.
In our subsequent analysis, we will use $k_\circ = 0.8$ and $k_\circ = 0.7$.

 \subsection{ Mercury with One Relaxation Time.}

 The generalized Voigt rheology (or standard solid rheology) with a single characteristic time
is represented by a spring-dashpot model shown in Figure \ref{genvoi} with $n=0$.
The Love number is described by equation (\ref{k2voigt2}) with $n=0$:
\begin{equation}
         k_2(\sigma) = k_\infty + (k_\circ - k_\infty)\,.
   \frac{1}{1 + i\tau\sigma}.\\  
        \label{k2voigt22}
      \end{equation}

\begin{figure}[hbt!]
\centering
\includegraphics[scale=0.6]{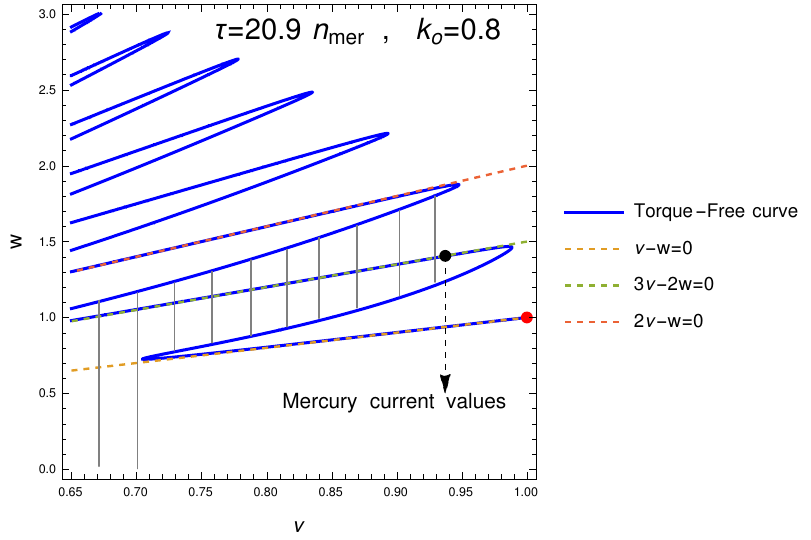}
\caption{Torque-free curve (blue) for Mercury with $k_\circ=0.8$ and
  one relaxation time: $\tau n_{merc} =20.91$. Three
spin-orbit resonance lines (dashed) are shown. Mercury is currently on the $3n = 2\omega$ resonance line. The red
point represents the synchronous equilibrium. The region hatched with vertical black lines
represents potential initial positions of Mercury that could have led to its current state.
The variables in the horizontal and vertical axes are:
$\nu= \frac{\ell^3_{\scriptscriptstyle T}}{a^3_1 } \, n$, ${\rm w}=
\frac{\ell^3_{\scriptscriptstyle T}}{a^3_1 } \, \omega$, where
$\frac{\ell^3_{\scriptscriptstyle T}}{a^3_1 }=\frac{0.9373}{n_{mer}}$.}
\label{t22}
\end{figure}

 In Figure \ref{t22}, we show the torque-free curve for the choice
 $k_\circ=0.8$. Imposing $k_2(n_{mer})=0.57-0.011\, i$ we obtain
 $k_\infty=0.5695$ and  $\tau n_{mer} = 20.91$.   Both the current
state of Mercury and the stable synchronous equilibrium are also depicted.
The current state of Mercury, as illustrated in Figure \ref{t22},
suggests that Mercury must have been captured
directly into the 3-2 spin-orbit resonance,
without prior transitions through higher-order resonances. 

\begin{figure}[hbt!]
\centering
\includegraphics[scale=0.6]{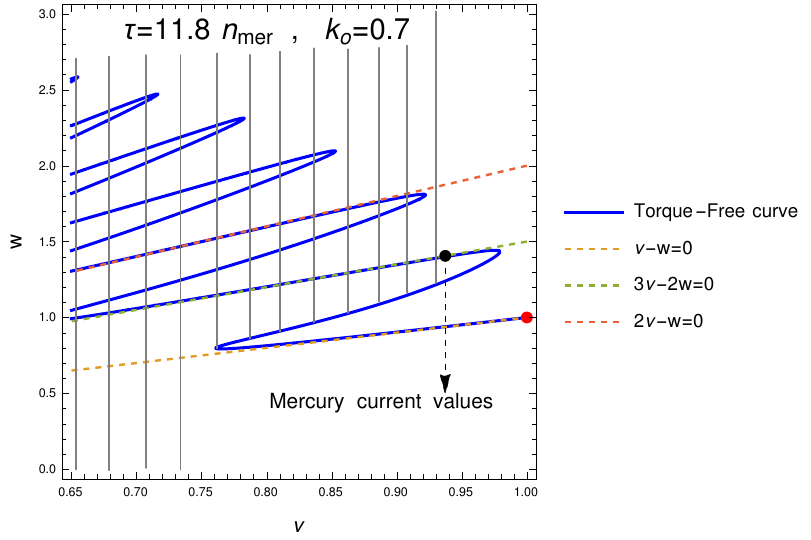}
\caption{Torque-free curve (blue) for Mercury with $k_\circ=0.7$ and one relaxation time $\tau n_{mer} = 11.82$. The remainder
of the notation is the same as that in Figure \ref{t22}.}
\label{t11}
\end{figure}

In Figure \ref{t11}, we show  the torque-free curve for the choice  $k_\circ=0.7$.
Imposing $k_2(n_{mer})=0.57-0.011\, i$ we obtain
 $k_\infty=0.5691$ and $\tau n_{mer} = 11.89$.
In this scenario, the current state of Mercury suggests that it could have been captured in
a higher-order resonance before settling into the 3-2 spin-orbit resonance.
The potential initial states of Mercury that lead to its current state,
as indicated by the region hatched with vertical lines in Figure \ref{t11},
encompass a much larger set than
the corresponding region in Figure \ref{t22}.

\subsection{Mercury with Two Relaxation Times.}

We  assume  a generalized Voigt  rheology, equation (\ref{k2voigt2}) and
Figure \ref{genvoi} with $n=1$.

We choose $k_\circ = 0.8$ and fix the complex Love number
at the orbital frequency as $k_2(n_{\text{mer}}) = 0.57 - 0.011\, i$.
Equation (\ref{k2voigt2}) then implies
 \begin{equation}
    0.57 - 0.011\,i = k_\infty + (0.8 - k_\infty)
         \left(\frac{h_1}{1 + i\tau_1 n_{\text{mer}}} + \frac{h_{2}}{1 + i\tau_{2}n_{\text{mer}}}\right)
           \label{eqk2}
 \end{equation}
with $h_1 > 0$, $h_2 > 0$, and
$h_1 + h_2 = 1$. This results in a set of two scalar equations for the four unknowns:
$k_\infty$, $0 < h_1 < 1$, $\tau_1 n_{\text{mer}}$, and $\tau_2 n_{\text{mer}}$.

In Section \ref{sols}, we demonstrate that equation (\ref{eqk2})
has solutions with arbitrarily large values
of $\frac{\tau_2}{\tau_1}$. According to the models discussed in Section \ref{terr},
the ratio between the largest and smallest relaxation times of both the Moon and Earth
is of the order of $10^5$. Therefore, it would be logical to choose solutions
for equation (\ref{eqk2}) where $\frac{\tau_2}{\tau_1}$ is at least of order $10^5$. However,
this choice presents numerical challenges in plotting the torque-free curve.
Consequently, some plots
presented below have a smaller  ratio of $\frac{\tau_2}{\tau_1}$. As argued
in Section \ref{sols}, the conclusions drawn would remain valid for any larger value of
$\frac{\tau_2}{\tau_1}$.

\begin{figure}[hbt!]
\centering
\includegraphics[scale=0.6]{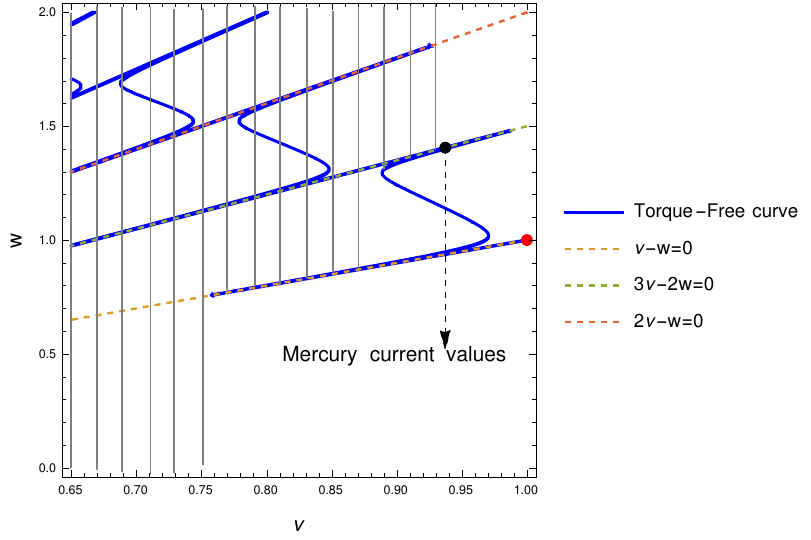}
\caption{Torque-free curve (blue) for Mercury with:
  $\tau_1 n_{mer}=0.48,  \tau_2 n_{mer}=1396, h_1= 0.11, h_2=0.89, k_\infty=0.5474$.
 The remainder
of the notation is the same as that in Figure \ref{t22}.}
\label{f1490}
\end{figure}

In Figure \ref{f1490}, we present the torque-free curve for the values:
$\tau_1 n_{\text{mer}} = 0.48, \tau_2 n_{\text{mer}} = 1396, h_1 = 0.11, k_\infty = 0.5474$, which
implies $k_2(n_{\text{mer}}) = 0.57 - 0.11\, i$.
In this scenario,
the current state of Mercury suggests that it could have initially been captured in
a higher-order resonance before transitioning to the 3-2 spin-orbit resonance.
The potential initial states of Mercury that could lead to its current state,
as indicated by the region hatched with vertical lines in Figure \ref{f1490},
encompass a much larger set than
the corresponding region in Figure \ref{t22}, which was derived using a rheology
with a single relaxation time under the same constraints
$k_2(n_{\text{mer}}) = 0.57 - 0.11\, i$ and $k_\circ=0.8$.

\begin{figure}[hptb!]
\centering
\begin{minipage}{0.5\textwidth}
\centering
\includegraphics[width=0.95\textwidth]{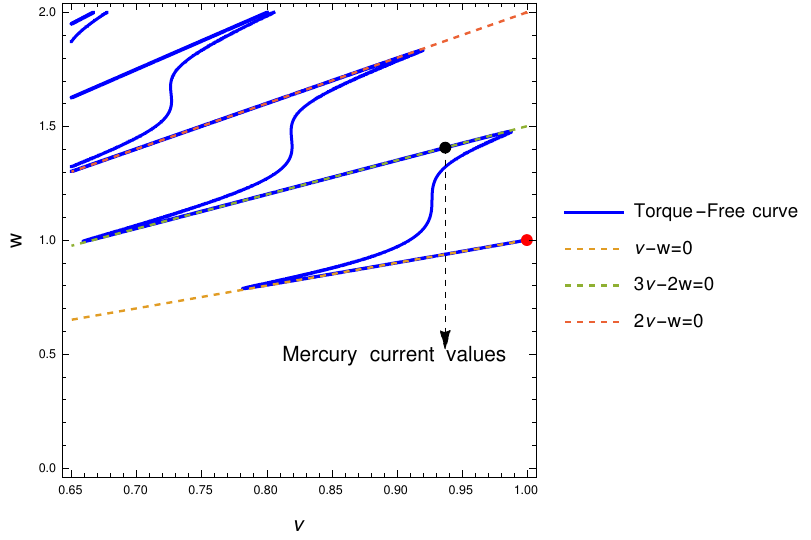}
\end{minipage}\hfill
\begin{minipage}{0.5\textwidth}
\centering
\includegraphics[width=0.95\textwidth]{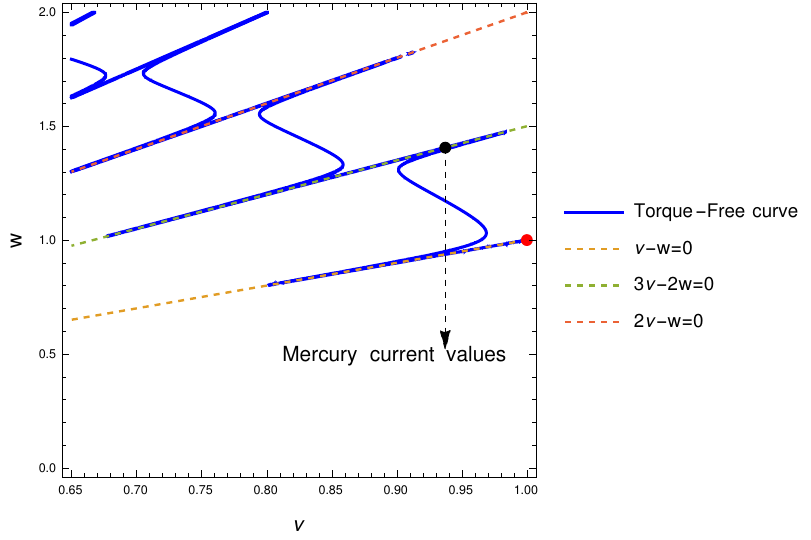}
\end{minipage}\hfill
\caption{Torque-free curve (blue) for Mercury.
   LEFT:
$\tau_1 n_{\text{mer}} = 0.015,  \tau_2 n_{\text{mer}} = 94, h_1 = 0.7126, k_\infty = 0$;
and  RIGHT:
$\tau_1 n_{\text{mer}} =0.019,   \tau_2 n_{\text{mer}} = 940, h_1 =0.7128, k_\infty = 0$.
 }
\label{f1000}
\end{figure}

In Figure \ref{f1000}, we illustrate the effect on the torque-free curve of increasing the ratio $\frac{\tau_2}{\tau_1}$ by a factor of 10. In both cases, the current state of Mercury suggests that it might have initially been captured in a higher-order resonance before transitioning to the 3-2 spin-orbit resonance. The conclusions regarding the set of initial conditions that lead to Mercury's current state, as established in Figure \ref{f1490}, are equally applicable to Figure \ref{f1000}.

The parameters used to generate Figure \ref{f1490} do not accurately reflect the current state of Mercury, where $\omega = \frac{3}{2} n_{\text{mer}}$. According to equation (\ref{amain3}) and with the parameters employed in Figure \ref{f1490}, the current rate of variation of Mercury's spin is $\frac{\dot \omega}{\omega} = \frac{1}{\Io \omega}\langle {\rm T}\rangle = -2.5 \times 10^{-7} \, \text{yr}^{-1}$. At this rate, over 10 million years, Mercury's spin would decrease to $10\%$ of its current value. The same conclusion applies to the parameters used to generate Figure \ref{f1000} and many other set of parameters that we tested. This may indicate that trapping into the $3:2$ spin-orbit resonance happened by means of a spin decrease and not the opposite.

\begin{figure}[hptb!]
\centering
\begin{minipage}{0.5\textwidth}
\centering
\includegraphics[width=0.95\textwidth]{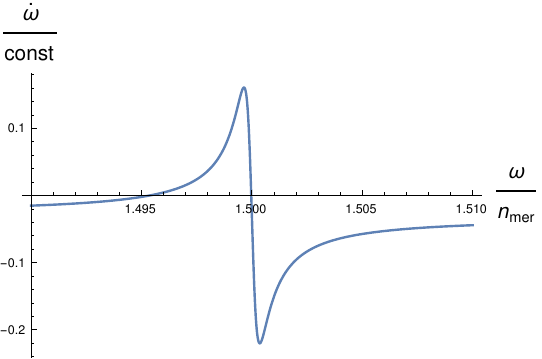}
\end{minipage}\hfill
\begin{minipage}{0.5\textwidth}
\centering
\includegraphics[width=0.95\textwidth]{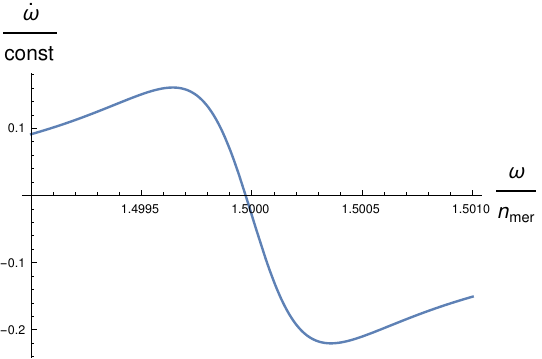}
\end{minipage}\hfill
\caption{Two graphs of $\dot\omega = \frac{\langle T\rangle}{\Io}$ divided by the constant 
$(k_\circ - k_\infty) \frac{3 n_{\text{mer}}^4 R^5}{2 G \Io} = 3.33 \times 10^{-19}\, \text{s}^{-2}$ 
are presented as a function of 
$\frac{\omega}{n_{\text{mer}}}$. The parameters are those used in Figure \ref{f1490}.
The graph on the left displays the two roots of $\dot \omega$, one of which appears to be at 
$\frac{\omega}{n_{\text{mer}}} = 1.5$. The graph on the right demonstrates that the root close to $\frac{\omega}{n_{\text{mer}}} = 1.5$ 
is actually slightly less than this value.
}
\label{torque}
\end{figure}

The strong decay rate of the spin seems paradoxical in light of Mercury's current position in Figure \ref{f1490}, which appears to be on the torque-free curve. This paradox is elucidated in Figure \ref{torque} left, where we can observe that the gradient of the torque function at a point on the torque-free curve is very large. Consequently, the torque may still be nonnegligible even at points quite close to the torque-free curve. In Figure \ref{torque} right, we demonstrate that for $n = n_{\text{mer}}$, the value of $\frac{\omega}{n_{\text{mer}}}$ where the torque is zero is slightly below $\frac{3}{2}$,
so Mercury is positioned above the torque-free curve.

\subsection{Conclusion}

In this paper, we have obtained, by means of energy arguments,
equations
for the secular evolution of spin and orbit due to tidal effects.
These equations were  originally obtained by other authors, notably in \cite{correia2022tidal}.
In the case of Mercury,
which served as a model in our study on the dynamical effects of increasingly complex rheologies,
the equations can be further simplified.

The main conclusion, which may be applicable to any binary system and not just Mercury-Sun,
is that employing
more complex and realistic rheologies is crucial in understanding spin-orbit evolution. 
This conclusion aligns with the findings in \cite{makarov2012conditions} and
\cite{noyelles2014spin}. According to the abstract of \cite{noyelles2014spin}:
``As opposed to the commonly used constant time lag (CTL) and constant phase lag (CPL) models, the
physics-based tidal model changes dramatically the statistics of the possible final spin states.''
Here, the term ``physics-based tidal model'' refers to a model derived from a rheology with
several relaxation times.

The past capture of Mercury into its current spin-orbit state has been the subject of
several previous studies. In our paper, we have completely neglected several important effects that may have been significant, as pointed out in \cite{noyelles2014spin}: perturbations by other bodies in the Solar System
that induce variations in Mercury's eccentricity, the potential relevance
of core-mantle-boundary friction at fluid-solid
interfaces, the possibility of past impacts with other bodies, existence of a permanent deformation,
and changes in the rheology
over long time scales. Within this broader context, two aspects of Mercury's rheology, which
were crucial in our model, 
may still remain significant: 1) The elastic rigidity
of Mercury at secular time scales, which
must exist; otherwise, Mercury would be in hydrostatic equilibrium. 2)
The presence of two or more relaxation times in Mercury's rheology.
Both these factors dramatically alter
the probability of Mercury being captured in its current state.
However, our quantitative results regarding the spin-orbit history of Mercury almost certainly 
do not align with what actually happened.

\section{Mathematical Complements}

In this section, we present several  mathematical facts that explain and clarify many
of the statements made in the previous section.

\subsection{The fundamental equations}

\label{eqsec}

 Let  $\kappa:=\{\mat e_1,\mat e_2,\mat e_3\}$ be an inertial  orthonormal frame
with origin at the centre of mass of the system. The orbit is contained in the plane
$\{\mat e_1,\mat e_2\}$.

Let  $\mathbf x$ be the vector from the center of mass of the small body to the  body $\alpha$.
The equation for the orbital motion is
\begin{equation}
  \begin{split}
  \ddot {\mathbf x} &= G(m_\alpha+m)  \bigg\{
- \frac{\mathbf x}{|\mathbf x|^3}
-\frac{15}{2}\frac{1}{|x|^7}\left( \left( \frac{\Ioa}{m_\alpha}  
\mathbf b_\alpha  + \frac{\Io}{m} \mathbf b \right)\mathbf x \cdot\mathbf x\right)\mathbf x  
  \\ & 
+3\frac{1}{|\mathbf x|^5}\left(  \frac{\Ioa}{m_\alpha}  \mathbf b_\alpha +
 \frac{\Io}{m} \mathbf b  \right)\mathbf x
\bigg\}\,,
\end{split}\label{xeq}
\end{equation}
where the deformation
matrix of the small body $\mathbf{b}$, defined in equation (\ref{Ib}), is given by
\begin{equation}
 \mat b=\left(
\begin{array}{ccc}
 b_{11} &  b_{12} & 0 \\
  b_{12} &  b_{22} & 0 \\
 0 & 0 &  b_{33} \\
\end{array}
\right)\,,\quad \text{with}\quad b_{33}=-b_{11}-b_{22}\,.
\label{B}
\end{equation}

For a rigid body, there exists a frame, namely the body frame, in which the body remains at rest.
Specifically, the angular momentum of the body with respect to the body frame is null. In the case of a deformable body, a frame still exists where the body's angular momentum is null: this is known as the Tisserand frame. Let $\K:=\{\mat e_{\sm{T1}},\mat e_{\sm{T2}}, \mat e_{\sm{T3}}\}$ denote an orthonormal Tisserand frame for the small body.
The orientation of $\K$ with respect to the inertial frame
$\kappa=\{\mat e_{1},\mat e_{2}, \mat e_{3}\}$ is described by $\mat R:\K\to\kappa$.
We assume that the angular velocity of the small
body, $\boldsymbol \omega$, is perpendicular to the orbital plane, which implies :
\begin{equation}
\mat R(\phi) = \left( \begin{array}{ccc} 
 \cos \phi  & -\sin \phi & 0 \\ 
 \sin \phi  & \cos \phi & 0 \\
 0 & 0 & 1
\end{array}  \right)\,,
\label{rot}
\end{equation}
with $\boldsymbol \omega=\omega \mat e_3=\dot\phi\mat e_3$.

The angular momentum of the small body is denoted by $ \boldsymbol{\ell}_s=\ell_s\mat e_3$, with the index $s$ representing spin, and is given by:
\begin{equation} \ell_s=\omega \Io(1-b_{3 3})\,.\label{eldef}
\end{equation}

In the context of the quadrupolar approximation, Euler's equation for the variation of $\ell_s$ is:
\begin{equation} \label{leq}
  \dot\ell_s=
  - \frac{3G \Io m_\alpha}{\|\mat x\|^5}\Big\{
x_1x_2 (b_{ 22}-b_{ 11})+b_{12} \left(x_1^2-x_2^2\right)\Big\}.
\end{equation}  
A similar equation holds for the large body.

  It is imperative to note that, unlike the case of a rigid body, the deformation matrix is not constant in the Tisserand frame $\K$.

  To complete the set of equations (\ref{xeq}) and (\ref{leq}), we require additional equations for
  the deformation matrices. These equations were derived within the Lagrangian formalism and
  utilizing what was termed the ``Association Principle'', as detailed in \cite{rr2015}, \cite{rr2017}.

  In the Tisserand frame $\K$ of the small body,
the deformation matrix and the position vector are denoted by capital letters as follows:
\begin{equation}
   \mathbf B=\mat R(\phi)\mat b\mat R^{-1}(\phi)\qquad \mat X= \mat R^{-1}(\phi)\mat x \,.
\end{equation}

The  equations for the deformation variables of the smaller body for the generalised Voigt model
in Figure \ref{genvoi}
are \cite[Eq. (3.15)]{ragazzo2022librations}
\begin{equation}
\begin{split}
&(\gamma+\alpha_0) \mathbf{B}+\mat \Lambda=\mathbf{F}\\
&\frac{1}{\alpha}\dot{\mat\Lambda}+\frac{1}{\eta} \mat\Lambda =\dot{\tilde{\mat B}}\\
&\eta_j\dot {\mathbf{B}}_{j}+\alpha_j\mat B_{j}=\mat\Lambda,\quad j=1,\ldots,n\\
&\mat B =\tilde{\mat B}+\mat  B_{1}+\mat  B_{2}+\ldots+\mat  B_{n}\, , 
\end{split}\label{maineq2}
\end{equation}
where:
\begin{itemize}
\item[$\bullet$] $\gm$, with dimensions of $1/$time$^2$, is a parameter representing the self-gravity rigidity of the body; a larger $\gamma$ indicates a stronger gravitational force holding the body together.
  In Figure \ref{genvoi} $\gm$ would be a spring in parallel to the spring-dashpot system
  representing the rheology.  
\item[$\bullet$] $\alpha_0$, $\alpha$, $\alpha_1,\ldots$ also with dimensions of $1/$time$^2$,
  are  elastic rigidity coefficients; $\alpha$ increases with the stiffness.
\item[$\bullet$] $\eta$, $\eta_1,\ldots$,  dimensions of  $1/$time, are viscosity parameters;
  a body with a larger $\eta$ is harder to deform at a given rate compared to a body with a smaller
  $\eta$.
\item[$\bullet$] $\tilde {\mat B}$, $\mat B_1,\ldots$,  are nondimensional traceless matrices
  that represent internal variables of the rheology;
\item[$\bullet$] $\mat \Lambda$, with dimensions 1/time$^2$, is a traceless  force matrix that
  represents the stress upon the rheology part ($\mat F-\mat \Lambda$ is the stress
  supported  by self-gravity).
\item[$\bullet$] $\mathbf{F}$, with dimensions 1/time$^2$, is the force matrix in the Tisserand frame
  $\K$:
\begin{equation}\renewcommand{\arraystretch}{1.5}
  \begin{array}{l l l}
    \mathbf{F}&:=\mat C+\mat S \quad&\text{Deformation force}\\ & & \\
    \mat C&:= \frac{\omega^2}{3}\left(\begin{matrix}
1& 0 & \ \ 0 \\ 0 & 1 & \ \ 0
\\ 0 & 0 & -2
\end{matrix}  \right)\quad&\text{Centrifugal force}\\ & & \\
     \mat S&:=
                   \frac{3G m_\alpha}{|\mat X|^5}\left(\mathbf{X}\mathbf{X}^T-
             \frac{|\mat X|^2}{3}\Id\right)\quad& \text{Tidal force}
\end{array}
\label{F}
\end{equation}
\end{itemize}
where $\mat X$ is undestood as a column vector, $\mat X^T$ is the transpose of $\mat X$ (a row vector
vector), and $\mat X\mat X^T$ is the usual product of a $n\times 1$ by $1\times n$   matrix, which
gives an $n\times n$ matrix with entries
$\big(\mat X\mat X^T\big)_{ij}=X_iX_j$. 
The norm of a matrix is defined as $\|\mat B\|^2=\frac{1}{2}\Tr \big(\mat B^T\mat B\big)$.

{\it Conservation of angular momentum.} 
The orbital angular momentum is given by
\begin{equation}
 \ell \mat e_3= \boldsymbol \ell=\frac{m_\alpha m}{m_\alpha+m} \mat x\times \dot{\mat x}\,.\label{ellvec}
\end{equation}
From  equations (\ref{xeq}) and (\ref{leq})  we obtain
\begin{equation}
  \dot{\boldsymbol \ell}= 3\frac{m_\alpha m}{|\mathbf x|^5}\mat x\times\left(  \frac{{\rm I}_{\circ\alpha}}{m_\alpha}  \mathbf B_\alpha +
    \frac{\Io}{m} \mathbf B  \right)\mathbf x=- \dot{\boldsymbol{\ell}}_{s\alpha}-
   \dot{\boldsymbol{\ell}}_s\,,\label{elleq}
  \end{equation}
that shows the conservation of  total angular momentum
  \begin{equation}
    \boldsymbol \ell_{\sm T}:=\boldsymbol \ell+\boldsymbol{\ell}_{s\alpha}+\boldsymbol{\ell}_s\,.\label{consl}
  \end{equation}
 
  {\it Energy.} 
 The total energy of the system is
  \begin{equation}E=E_{kcm}+E_{krot}+E_{def}+E_{gr}\label{etotal}\end{equation}
  where:
  \begin{equation}
    E_{kcm}=\frac{m_\alpha m}{m_\alpha+m}\frac{|\dot{\mathbf x}|^2}{2}
  \end{equation}
  is the kinectic  energy  of the orbital motion;
  \begin{equation}
    E_{krot}=\frac{\boldsymbol{\omega}_\alpha\cdot \mathbf{I}_\alpha\boldsymbol{\omega}_\alpha}{2}
    +\frac{\boldsymbol{\omega}\cdot \mathbf{I}\boldsymbol{\omega}}{2}
   \end{equation}
is the kinectic energy of rotation;
\begin{equation}
  E_{def}=
     \frac{\Io}{2}\left((\gm+\alpha_0) \|\mathbf{B}\|^2 + \frac{\|\mat\Lambda\|^2}{\alpha}
+\sum_{j=1}^n\alpha_{ j}\|\mat B_{j}\|^2\right)+\ \text{Large body term}
\end{equation}
is the elastic energy of deformation of the generalised Voigt  model in Figure \ref{genvoi}; and
\begin{equation}\begin{split}
  E_{gr}=&
   -
\frac{G m_\alpha m}{|\mathbf x|}
-\frac{3G}{2}\frac{1}{|\mathbf x|^5}
\bigl\{m{\rm I}_{\circ \alpha} \big(\mathbf x\cdot \mathbf{B}_\alpha
\mathbf x\big)+m_\alpha\Io( \mathbf x\cdot \mathbf{B}\mathbf x)\bigr\}
\end{split}
\end{equation}
is the  potential energy due to gravitational interactions.

{\it Dissipation of energy.} For  the generalised Voigt  model in Figure \ref{genvoi},
the dissipation function of the small body is:
 \begin{equation}
   \mathcal{D}= 
   \frac{\Io}{2}\left(\frac{\|\mat\Lambda\|^2}{ \eta}+  \sum_{j=1}^n\eta_j\|\dot{\mat B}_j\|^2\right)
   \label{D}
\end{equation}
A similar expression $\mathcal{D}_\alpha$  holds for the large body.
The Lagrangian formalism with dissipation
function \cite{rr2015}, \cite{rr2017} implies that the total energy decreases along the motion: 
\begin{equation}
    \dot E=-(2\mathcal D+2\mathcal D_\alpha)\le 0\,.
  \end{equation}
  Moreover, $2\mathcal D_\alpha$ and $2\mathcal D$ are the powers dissipated within the large and small
  bodies, respectively. 

  The secular equations (\ref{amain})  are  approximations to the
  fundamental equations presented in this Section.

\subsection{Computation of the Passive Deformation Matrix.}

\label{passec}

Passive deformation refers to the tides caused by an orbiting point mass on an extended body,
while the influence of the gravitational field, resulting from the deformation on the orbit, is neglected.
The equations that describe these passive deformations are obtained setting $\mat B=\mat B_\alpha=0$
in equations (\ref{xeq}) and (\ref{leq}) while preserving equation (\ref{maineq2}):
\begin{equation}
    \begin{split}  
      & \ddot {\mathbf x} =- G(m_\alpha+m)  \frac{\mathbf x}{|\mathbf x|^3}\,,\quad  \dot \omega =0\,,
      \quad \dot \omega_\alpha =0\\
  &(\gamma+\alpha_0) \mathbf{B}+\mat \Lambda=\mathbf{F}\\
&\frac{1}{\alpha}\dot{\mat\Lambda}+\frac{1}{\eta} \mat\Lambda =\dot{\tilde{\mat B}}\\
&\eta_j\dot {\mathbf{B}}_{j}+\alpha_j\mat B_{j}=\mat\Lambda,\quad j=1,\ldots,n\\
&\mat B_{T} =\tilde{\mat B}+\mat  B_{1}+\mat  B_{2}+\ldots+\mat  B_{n}\,.\\
   &\text{Plus deformation equations for the large body}
\end{split}\label{epd2}
\end{equation}
In this scenario $\omega$ and $\omega_\alpha$  remain constant and $\mat x$ follows a Keplerian ellipse.

In order to write the $\mat F$  in a convenient way, we define the matrices  \cite[Eq. (41)]{rr2017}
 {\small
 \begin{equation}
\mathbf{Y}_0:=
\frac{1}{\sqrt{3}}\left(\begin{matrix}
1& 0 & \ \ 0 \\ 0 & 1 & \ \ 0
\\ 0 & 0 & -2
\end{matrix}  \right)\,,\quad
\mathbf{Y}_{1}:=
\frac{1}{\sqrt{2}}
 \left(\begin{matrix}
0 & 0 & 1 \\ 0 & 0 & -i 
\\ 1 & -i & 0
\end{matrix}  \right)\,,\quad 
\mathbf{Y}_2:=
\frac{1}{\sqrt{2}}\left(\begin{matrix}
1 & -i & 0 \\ -i & -1 & 0
\\ 0 & 0 & 0
\end{matrix}  \right)\,,
\label{Yj}
\end{equation}}
$\mat Y_{-1}=\ov{\mat Y}_1$,  and $\mat Y_{-2}=\ov{\mat Y}_2$, where the overline  represents
complex conjugation.

Any symmetric matrix can be writen as a linear combination of the six matrices
$\{\Id,\mat Y_{-2},\mat Y_{-1},\mat Y_0,\mat Y_1,\mat Y_2\}$. This basis is orthonormal
with respect to the Hermitean inner product  $\mathbf{A}\cdot\mathbf{B} =\frac{1}{2}
  \Tr\big(\, \ov{\mathbf{A}}^{\rm T}\mathbf{B}\big)=
  \frac{1}{2}\sum_{ij}\ov A_{ij}B_{ij}$.

  Traceless matrices that are invariant under rotation  with respect to the $\mat e_3$-axis can
  be written as linear combinations of $\{\mat Y_{-2},\mat Y_0,\mat Y_2\}$.
These matrices have a simple transformation rule with respect to  rotations about the axis
$\mat e_3$, namely
  \begin{equation}
    \mat R(\theta)\mat Y_j\mat R^{-1}(\theta)=\erm^{i\, j\, \theta} \,\mat Y_j\,,\quad j=-2,0,2\,.
\label{eig}    \end{equation}

An elliptic orbit in the body frame is given by
\begin{equation}\begin{split}
    \mathbf X&=\mat R^{-1}(\phi)\mat x=  r \mat R(f+\varpi-\phi)\mat e_1\\
    &=
    r(\cos (f+\varpi-\phi)\,  \mat e_1+\sin (f+\varpi-\phi)\, \mat e_2)\,,
    \end{split}
\end{equation}
where $f$ is the mean anomaly, $\varpi$ is the argument of the periapsis.

The associated tidal-force matrix,  equation (\ref{F}), can be written as  
\begin{equation}\begin{split}
  \mat S&=\frac{3 G m_\alpha}{r^3} \mat R_3(f+\varpi-\phi)\Big\{\mat e_1\mat e_1^T-  \frac{1}{3}\Id\Big\} \mat R^{-1}_3(f+\varpi-\phi)\\ & =\frac{3 G m_\alpha}{2r^3}
 \bigg\{\erm^{-2 i (f+\varpi-\phi)}\frac{ \mat Y_{-2}}{\sqrt 2}+\frac{\mat Y_0}{\sqrt 3}+
  \erm^{2 i (f+\varpi-\phi)}\frac{\mat Y_2}{\sqrt 2}\bigg\}. \end{split}\label{tf}
  \end{equation}
              In equation (\ref{tf}), the variables $r$, $f$, and $\phi=\omega t$ are dependent on $t$.

Equations (\ref{tf}) and (\ref{hanseneq}) gives a harmonic decomposition of the tidal force as follows:
\begin{equation}
 \mat S=
    \frac{3G m_\alpha}{2a^3} \sum_{l=-2}^2\sum_{k=-\infty}^\infty
   \erm^{i\{t(k n-l\omega)+ l\,\varpi\}}
   \mat Y_lU_{kl},\label{Saap2}
\end{equation}
where $U_{k,-1}=U_{k,1}=0$ and 
\begin{equation}
  U_{k,-2}=\frac{X^{-3,-2}_k}{\sqrt 2}\,,  \ \ 
 U_{k0}=\frac{X^{-3,0}_k}{\sqrt 3}\,,  \ \ 
 U_{k2}=\frac{ X^{-3,2}_k}{\sqrt 2} \,.\label{Uk02}
\end{equation}
The symmetry property $X^{n,-m}_{-k} = X^{n,m}_{k}$ implies $ U_{kj}=U_{-k,-j}$.

The centrifugal force in equation (\ref{F}) can be represented as
\begin{equation}
  \mat C = \frac{\omega^2}{\sqrt 3}\mat Y_0\,. \label{C2}
\end{equation}

To obtain 
 the almost periodic solution of the deformation equation
 solving for each Fourier mode separately suffices.

To simplify the notation we consider a simple harmonic force term of the form
\[ \mat F(t)=\mat {F}'\erm^{i\, \sigma\, t}\]
where $\mat {F}'$ is a complex amplitude matrix,
and $\sigma\in\R$ is the constant forcing frequency.

If we do the substitutions 
\[
 \mat B\to \mat {B}'\, \erm^{i\, \sigma \, t}\,,\quad \mat B_j\to \mat {B}_j'\, \erm^{i\, \sigma \, t}\,,\quad
 \mat \Lambda\to \mat {\Lambda}'\,\erm^{i\, \sigma \, t}\,,
\]
 where $\mat {B}'$,  $\mat {B}_j'$ and  $ \mat {\Lambda_j}'$
are understood as
constant complex-amplitude matrices, into  equation     (\ref{maineq2}), then we obtain
after some simplifications
\begin{equation}
\left\{ \gamma+\alpha_0+\left(\frac{1}{\alpha}+  \frac{1}{\eta i\sigma}+
  \sum_{j=1}^n\frac{1}{\alpha_j+i\sigma \eta_j}\right)^{-1}\right\}\mat B'=
\mat F'\,.
   \label{love1.1}
  \end{equation}
 The term $\frac{1}{\alpha}+  \frac{1}{\eta i\sigma}+
 \sum_{j=1}^n\frac{1}{\alpha_j+i\sigma \eta_j}$ 
 is the complex  rigidity  $J(\sigma)$ 
 of the generalised Voigt  model when $\alpha_0=0$
 \cite[equation 4.23]{ragazzo2022librations}.
 
 The complex Love number at frequency $\sigma$ can be written as
 \cite[Section 4]{correia2018effects}:
\begin{equation}
  k_2(\sigma)=
  \frac{3\Io G}{R^5}\frac{1}{\gamma+\alpha_0+J^{-1}(\sigma)}\,.
\label{Lovegenvoigt}
\end{equation}
Note: $k_2(-\sigma)=\ov k_2(\sigma)$.
Therefore, equation (\ref{love1.1}) can be written as
\begin{equation}
  \mat B'=  \frac{R^5}{3\Io G}k_2(\sigma)\mat F'.\label{love1.2}
\end{equation}

Applying equation (\ref{love1.2}) to each Fourier coefficient of $\mat F = \mat S + \mat C$,
where $\mat S$ and $\mat C$ are defined in equations (\ref{Saap2}) and (\ref{C2}) respectively,
we obtain the passive-deformation matrix
  \begin{equation}\begin{split}
  \mat B(t)&=k_\circ \zeta_c
 \frac{\mat Y_0}{3\sqrt 3}+
  \zeta_{\scriptscriptstyle T}\sum_{l=-2}^2\sum_{k=-\infty}^\infty
   \erm^{i\{t(k n-l\omega)+ l\,\varpi\}}
   k_2(k n-l\omega)  \mat Y_lU_{kl},
 \end{split}
 \label{DFWb}
 \end{equation}
 where $ k_\circ=k_2(0)$ is the secular Love number and
 \begin{equation}
  \zeta_c := \frac{R^5 \omega^2}{G\Io} \quad \text{and} \quad
  \zeta_{\scriptscriptstyle T} := \frac{m_\alpha R^5}{2\Io a^3}.
\label{zeta}
\end{equation}

  \subsection{The Average Dissipation of Energy and Average Torque.}
  \label{secav}

The energy dissipated in the smaller body is $-2\mathcal{D}$, where $\mathcal{D}$ is the dissipation
function as given in equation (\ref{D}). The average dissipation function over the orbital motion 
is approximately given by 
\begin{equation}
  \langle \mathcal{D}\rangle := \lim_{t\to\infty}\int\limits_0^T\mathcal{D}dt
 = \frac{\Io}{2}\lim_{t\to\infty}\int\limits_0^T
 \left(\frac{\|\mat\Lambda\|^2}{ \eta}+  \sum_{j=1}^n\eta_j\|\dot{\mat B}_j\|^2\right)dt\,,\label{Dav}
\end{equation}
where $(\mat B, \mat \Lambda, \ldots)$ are the components of the solution to equation
(\ref{maineq2}) under the passive deformation hypothesis.

From equation  (\ref{maineq2})  we obtain:
\begin{equation}
\begin{split}
  &(\gamma+\alpha_0)\bcancel{\langle\dot {\mat B}\cdot \mathbf{B}\rangle}+\langle\dot {\mat B}\cdot\mat \Lambda
  \rangle=\bcancel{\langle\dot {\mat B}\cdot\mathbf{C}\rangle}+\langle\dot {\mat B}\cdot\mathbf{S}\rangle\\
  &\bcancel{\frac{1}{\alpha}\langle\mat\Lambda\cdot\dot{\mat\Lambda}\rangle}+
  \frac{1}{\eta} \|\mat\Lambda\|^2 =\langle\mat \Lambda\cdot\dot{\tilde{\mat B}}\rangle\\
  &\eta_j\|\dot {\mathbf{B}}_{j}\|^2+\alpha_j\bcancel{\langle\dot{\mat B}_j\cdot\mat B_{j}\rangle}=
  \langle\dot{\mat B}_j\cdot\mat\Lambda\rangle,\quad j=1,\ldots,n\\
  &\langle\dot{\mat B}\cdot\mat \Lambda\rangle =\langle\dot{\tilde{\mat B}}\cdot\mat \Lambda\rangle+
  \langle\dot{\mat  B}_{1}\cdot\mat \Lambda\rangle+\langle\dot{\mat  B}_{2}\cdot\mat \Lambda\rangle+\ldots+
  \langle\dot{\mat  B}_{n}\cdot\mat \Lambda\rangle\,,
\end{split}\label{maineq22}
\end{equation}
where we used that  the centrifugal force $\mat C$, equation (\ref{C2}), is constant.
These equations imply
\begin{equation}
  \langle \mathcal{D}\rangle =
 \frac{\Io}{2}\langle\dot{\mathbf{B}}\cdot\mat S\rangle\,,\label{Dav2}
\end{equation}
where $\mat S$ and  $\mathbf{B}$ are given in equations  (\ref{Saap2}) and (\ref{DFWb}),
respectively.

In the computation of the averaging $\langle\dot{\mathbf{B}}\cdot\mat S\rangle$ the following
identities are useful: $\mat Y_l\cdot \mat Y_{-m}=\delta_{lm}$, $U_{kl}=U_{-k,-l}$,
and $k_2(-\sigma)=\ov k_2(\sigma)$. After some computations  we obtain 
equation (\ref{da1}).

The total energy, as given in equation (\ref{etotal}), includes terms that depend on the deformation
matrices. If we set $\mat B = \mat B_\alpha = 0$ in this equation and assume
that the orbit is Keplerian and the spins are constant, then we obtain
the energy function $E_\circ$, as presented in equation (\ref{Eo}), for a system of two rigid spherical bodies.
It is a natural assumption that the energy dissipated in the two bodies is subtracted from $E_\circ$.
This assumption leads to equation (\ref{th1}).

The average torque is computed as the average of the right-hand side of equation (\ref{leq}),
\begin{equation}
  \langle {\rm T}\rangle := \lim_{t\to\infty}\int\limits_0^T
  - \frac{3G \Io m_\alpha}{\|\mat x\|^5}\Big\{
x_1x_2 (b_{22} - b_{11}) + b_{12} \left(x_1^2 - x_2^2\right)\Big\} dt
\,,\label{Tav}
\end{equation}
where $\mat B$ represents the passive deformation matrix and $\mat x$ determines Keplerian ellipses.
Standard computations lead to equation (\ref{avt1}).

Equations (\ref{th1}) and (\ref{avt1}) imply equations  (\ref{amain}).

\subsection{Solutions to Equation (\ref{eqk2}) for $\tau_2 n_{\text{mer}} \gg 1$ and the Torque-Free Curve for $\tau_2/\tau_1 \gg 1$}
\label{sols}

\begin{figure}[hbt!]
\centering
\includegraphics[scale=0.6]{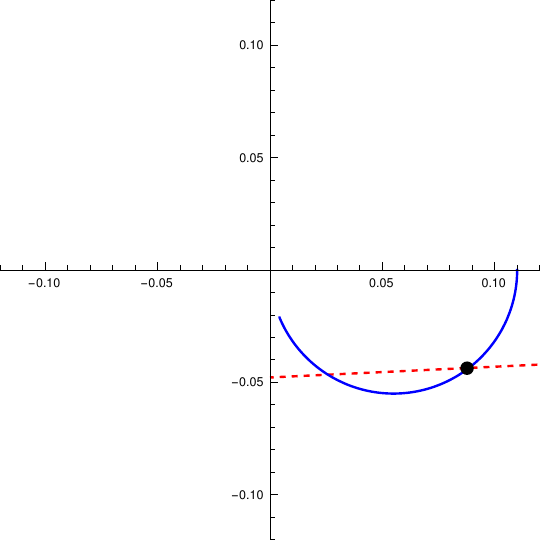}
\caption{Graphical solution to equation (\ref{eqk22}) in the limit as
$\tau_2 \to \infty$ for $h_1 = 0.11$. The dashed-red curve is parameterized by
$k_\infty$, while the blue curve is parameterized by $\tau_1$. The intersections of these curves represent the solutions
$(k_\infty, \tau_1 n_{\text{mer}})$, specifically: $(0.547658, 0.492353)$, which is marked by a ball, and
$(0.563915, 1.80765)$.
}
\label{int1a}
\end{figure}

If $\tau_2 n_{\text{mer}} \gg 1$, then equation (\ref{eqk2}) can be approximately written as
 \begin{equation}
    \frac{0.57 - k_\infty}{0.8 - k_\infty} - \frac{0.011}{0.8 - k_\infty}\,i =  
       \frac{h_1}{1 + i\tau_1 n_{\text{mer}}} - i \frac{1 - h_1}{\tau_{2}n_{\text{mer}}}.
           \label{eqk22}
 \end{equation}
 The left-hand side defines a curve in the complex plane, parameterized by
 $k_\infty \in [0, 0.57]$. As $\tau_2 \to \infty$, the right-hand side converges to
 $\frac{h_1}{1 + i\tau_1 n_{\text{mer}}}$, and for a given $h_1 \in (0, 1)$, it defines a curve
 in the complex plane, parameterized by $\tau_1 \in (0, \infty)$. This curve is a semi-circle
 below the real axis, with a radius of $\frac{h_1}{2}$ and centered at $\left(\frac{h_1}{2}, 0\right)$.
In the limit as
 $\tau_2 \to \infty$, for a given $h_1$, the solution to equation (\ref{eqk22}) is represented by a pair
 $(k_\infty, \tau_1)$, corresponding to the intersection of
 these two curves, as illustrated in Figure \ref{int1a}.

In the following, we graphically demonstrate
how to obtain a solution to equation (\ref{eqk2}) for $\tau_2 n_{\text{mer}} \gg 1$, 
using the solution marked with a ball in Figure \ref{int1a}.

\begin{figure}[hbt!]
\centering
\includegraphics[scale=0.5]{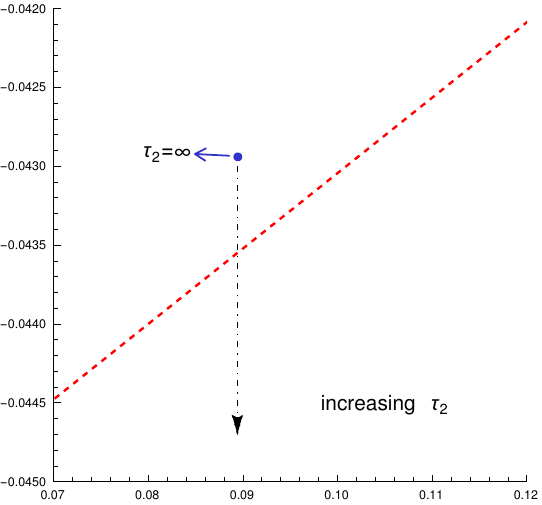}
\caption{Graphical solution to equation (\ref{eqk23}) for
  $\tau_2 n_{mer}\gg 1$ and  $h_1=0.11$. The dashed-dotted-black line is parameterized by
  $\tau_2$.}
\label{int2a}
\end{figure}

For the fixed value of $h_1 = 0.11$, we choose $\tau_1 n_{\text{mer}}$ to be slightly smaller
than $0.492353$, which corresponds to the solution marked 
in Figure \ref{int1a}, say $\tau_1 n_{\text{mer}} = 0.48$.
The curve parameterized by $\tau_1 \in [0, 0.48/n_{\text{mer}}]$ (blue) will terminate slightly above
the curve parameterized by $k_\infty$ (red).
Substituting $h_1 = 0.11$ and $\tau_1 n_{\text{mer}} = 0.48$ into equation (\ref{eqk22}), we obtain 
 \begin{equation}
    \frac{0.57 - k_\infty}{0.8 - k_\infty} - \frac{0.011}{0.8 - k_\infty}\,i =  
       \frac{0.11}{1 + i 0.48} - i \frac{0.89}{\tau_{2}n_{\text{mer}}}.
           \label{eqk23}
 \end{equation}
Now, the right-hand side of equation (\ref{eqk23}) defines a vertical segment in the complex plane
starting at $\frac{0.11}{1 + i 0.48}$ for $\tau_2 = +\infty$, and decreasing to $- \infty \, i$
as $\tau_2 \to 0$. This curve clearly intersects the curve parameterized by $k_\infty$, as depicted
in Figure \ref{int2a}. As $\tau_1 n_{\text{mer}}$ approaches $0.492353$ from below, the value of
$\tau_2 n_{\text{mer}}$ increases towards $+\infty$.

An application of the implicit function theorem to equation (\ref{eqk22}) around the solution
$(k_\infty, \tau_1 n_{\text{mer}}, (\tau_2 n_{\text{mer}})^{-1}) = (0.547658, 0.492353, 0)$ yields the following formula:
 \begin{equation}
   k_\infty= 0.547658 -\frac{0.341496}{\tau_2 n_{mer}}\,,\quad
   \tau_1 n_{mer}= 0.492353 -\frac{17.5779}{\tau_2 n_{mer}}. \label{tauap}
   \end{equation}

As previously mentioned, plotting the torque-free curve is numerically challenging when $\tau_1 \ll \tau_2$. In this limit, it is possible to approximate the torque-free curve near a given resonance line, $k_*n - 2\omega = 0$, as demonstrated in the following paragraphs.

First, we substitute the variable $\omega$ with $\epsilon$ using the equation $k_*n - 2\omega = -2\epsilon$. The analysis will be confined to a neighborhood of the resonance line, namely
\[
  |\epsilon|\tau_2 < 1\,.
\]

We write  equation (\ref{Ckv}), for the torque-free curve,  using the new variable
\[\begin{split}
 0&=-\frac{1}{k_\circ-k_\infty}\sum\limits_{k=-\infty}^\infty
       \Big(X^{-3,2}_{k}(e)\Big)^2{\rm Im} \, k_2\big((k-k_*)n-2\epsilon\big)=
      \\ &  
 \sum\limits_{k=-\infty}^\infty
       \Big(X^{-3,2}_{k}(e)\Big)^2
       \left(\frac{h_1\tau_1\big((k-k_*)n-2\epsilon\big)}{1+\tau_1^2\big((k-k_*)n-2\epsilon\big)^2}
 + \frac{(1-h_1)\tau_2\big((k-k_*)n-2\epsilon\big)}{1+\tau_2^2\big((k-k_*)n-2\epsilon\big)^2}\right).
  \end{split}     
\]
For $k \neq k_*$, where $|\pm n - 2\epsilon| > \text{constant} > 0$, and $\tau_2 \gg 1$, 
\[
  \left|\frac{\tau_2\big((k - k_*)n - 2\epsilon\big)}{1 + \tau_2^2\big((k - k_*)n - 2\epsilon\big)^2}\right| <
   \left|\frac{1}{\tau_2\big((k - k_*)n - 2\epsilon\big)}\right| \approx 0\,,
\]
and the equation for the torque-free curve can be approximately written as
\[
 \sum\limits_{k=-\infty}^\infty
       \frac{\Big(X^{-3,2}_{k}(e)\Big)^2}{\Big(X^{-3,2}_{k_*}(e)\Big)^2}
       \frac{\tau_1\big((k - k_*)n - 2\epsilon\big)}{1 + \tau_1^2\big((k - k_*)n - 2\epsilon\big)^2}
        =  \frac{1 - h_1}{h_1}\frac{2\tau_2\epsilon}{1 + (2\tau_2\epsilon)^2}\,,
\]
where, according to approximation (\ref{en}),
$e = \sqrt{1 - \left(\frac{\ell^3_{\scriptscriptstyle T}}{a^3_1 } n\right)^{2/3}}$.
It is convenient to rewrite this equation using the variables
$\nu := \frac{\ell^3_{\scriptscriptstyle T}}{a^3_1 } \, n$, 
${\rm w} := \frac{\ell^3_{\scriptscriptstyle T}}{a^3_1 } \, \omega$,
$T_1 := \frac{a^3_1 }{\ell^3_{\scriptscriptstyle T}}\tau_1$, and $x = 2\epsilon \tau_2$:
\begin{equation}
\sum\limits_{k=-\infty}^\infty
       \frac{\Big(X^{-3,2}_{k}(e)\Big)^2}{\Big(X^{-3,2}_{k_*}(e)\Big)^2}
       \frac{\big((k - k_*)T_1 \nu - \frac{\tau_1}{\tau_2}x\big)}{1 + \big((k - k_*)T_1\nu - 
       \frac{\tau_1}{\tau_2}x \big)^2}
        =  \frac{1 - h_1}{h_1}\frac{x}{1 + x^2}\,, \label{fau}
\end{equation}
where
$e = \sqrt{1 - \nu^{2/3}}$.

We will assume that $0.65 < \nu \leq 1$. There are two reasons for this assumption. Firstly, we computed Hansen's coefficients up to the 20th order in $e$, which imposes the limitation $\nu > 0.65$. The second reason is that we will assume $\frac{\tau_1}{\tau_2} \ll T_1 \nu$. 
This inequality, combined with $|\epsilon| \tau_2 < 1$ and $|x| < 2$, implies that for $k \neq k_*$, $(k - k_*)T_1 \nu - \frac{\tau_1}{\tau_2}x \approx (k - k_*)T_1 \nu$.
Therefore, equation (\ref{fau}) can be approximated by
\begin{equation}
\sum\limits_{k \neq k_*}
       \frac{\Big(X^{-3,2}_{k}(e)\Big)^2}{\Big(X^{-3,2}_{k_*}(e)\Big)^2}
       \frac{\big((k - k_*)T_1 \nu \big)}{1 + \big((k - k_*)T_1 \nu \big)^2}
        =  \frac{1 - h_1}{h_1}\frac{x}{1 + x^2}\,,\label{fau2}
\end{equation}  
where the term $\frac{\frac{\tau_1}{\tau_2}x}{1 + \big(\frac{\tau_1}{\tau_2}x \big)^2}$,
corresponding to $k = k_*$, was neglected because $\left|\frac{\tau_1}{\tau_2}x\right|$ is small.

\begin{figure}[hbt!]
\centering
\includegraphics[scale=0.7]{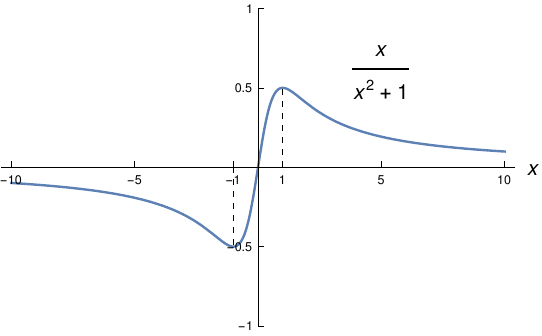}
\caption{The function on the right-hand side of equation  (\ref{fau2}).}
\label{plot}
\end{figure}

\begin{figure}[hbt!]
\centering
\includegraphics[scale=0.7]{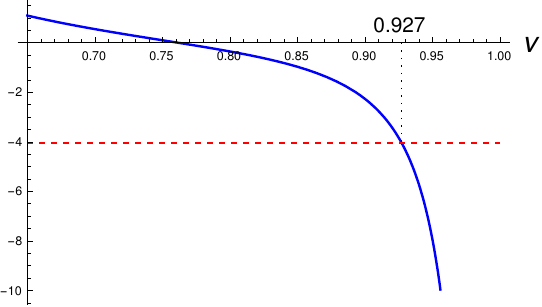}
\caption{Graph (solid blue curve)  of the function  on the left-hand side of equation  (\ref{fau2}) for
  $T_1=0.512$, which corresponds to $\tau_1 n_{mer}=0.48$. The horizontal line (brown-dashed)
corresponds to $-4.045=-\frac{1-h_1}{2h_1}$, $h_1=0.11$.}
\label{left}
\end{figure}

The function on the right-hand side of equation (\ref{fau2}) depends solely on $x$
and, except for the factor $\frac{1 - h_1}{h_1}$, it is plotted in Figure \ref{plot}. The
function on the left-hand side of equation (\ref{fau2}) depends only on $\nu$ and $T_1$.
For $k_* = 4$, which corresponds to the resonance $2n - \omega = 0$, this graph
with $T_1 = 0.512$ is plotted in Figure \ref{left}.

The function on the right-hand side of equation (\ref{fau2}) has its minimum  
equal to $-\frac{1 - h_1}{2h_1}$ and its maximum equal to $\frac{1 - h_1}{2h_1}$.
For a given value of $h_1 \in (0,1)$, the maximum value
of $\nu$ on the torque-free curve
is determined by the intersection of the horizontal line $-\frac{1 - h_1}{2h_1}$ and the
graph of the function on the left-hand side of equation (\ref{fau2}), as depicted
in Figure \ref{left}. This result should be accurate in the limit as $\tau_2 \to +\infty$ and
is likely reasonable for $\tau_2/\tau_1 \gg 1$. 

Using Figure \ref{left}, we estimated that on the torque-free curve, close to 
the resonance $2n - \omega = 0$, and for $T_1 = 0.512$ (which corresponds
to $\tau_1 n_{\text{mer}} = 0.48$), $\nu$ attains a maximum value of $0.927077$.
This value is close to $0.9267$, which is the value for the corresponding point in Figure \ref{f1490},
where $\frac{\tau_2}{\tau_1} = 2900$.

\section*{Acknowledgements}
 L. R. S.  is supported in part by FAPEMIG (Funda\c{c}\~ao de Amparo \`a Pesquisa no
Estado de Minas Gerais) under Grants No. RED-00133-21 and APQ-02153-23.

\bibliographystyle{plainnat}
\bibliography{mybibliography}

\end{document}